\documentclass[12pt,showpacs,preprintnumbers,amsmath,aps,amssymb,epsfig]{revtex4-1}
\usepackage{graphicx}
\usepackage{dcolumn}
\usepackage{bm}
\usepackage{color}

\begin{document}
\title{Comments On MacDowell-Mansouri gravity, Torsion  and $\Lambda$.}
\author{J. E. Rosales-Quintero}
\email{erosales@fisica.ugto.mx}
\author{M. Sabido}
\email{msabido@fisica.ugto.mx}
\affiliation{Departamento de F\'isica, DCI, Campus Le\'on, Universidad
  de Guanajuato, A.P. E-143, C.P. 37150, Le\'on, Guanajuato, M\'exico.}
\altaffiliation[Present Address:]{Department of Theoretical Physics, University of the Basque Country UPV/EHU, P.O. Box 644, 48080 Bilbao, Spain.}
\author{J. C. L\'opez-Dom\'inguez}
\email{jlopez@fisica.uaz.edu.mx}
\affiliation{Unidad Ac\'ademica de F\'isica\\
 Universidad Aut\'onoma de Zacatecas, Calzada Solidaridad esquina con Paseo a la Bufa S/N, C.P. 98060, Zacatecas, Zacatecas, M\'exico.}
\begin{abstract}
Starting with the MacDowell-Mansouri
formulation of gravity with a $SO(4,1)$ gauge group, we introduce new
parameters into the action to include the
non-dynamical Holst term, and the topological
Nieh-Yan and Pontryagin classes. Then, we consider the new parameters as fields
and analyze the solutions coming from their equations of motion.
The new fields introduce torsional contributions to the theory
that modify Einstein's equations. 
\end{abstract}
\maketitle

\section{Introduction}

The success of Loop Quantum Gravity (LQG) can be traced to the
polynominal nature of Ashtekar formulation, this approach is
independent of background structures of space time. It was pointed
out by Barbero \cite{barbero}, that a formulation of gravity based on  a
pure real connection can be constructed, but  the
constraints in Barbero's Hamiltonian formulation are more
complicated (this is the price to pay for a real formulation). 
Immirzi \cite{Immirzi}, noted that Barbero's transformations
can be generalized to a one-parameter family of transformations. 
Classically the Barbero-Immirzi (BI) parameter $\gamma$
appears as a free parameter with no physical meaning, but is
of significant relevance  at the quantum level. {A generalized
Hilbert-Palatini (HP) action containing the BI parameter was proposed \cite{Holst}}.
{It has been suggested that the} BI parameter has similarities to the $\theta$-ambiguity  that arises in Yang-Mills theories  \cite{obregon}.
The MacDowell-Mansouri (MM) action \cite{MacDowell-Mansouri} with a topological $\theta$-term, offered a possibility to interpret the BI parameter as the $\theta$-parameter \cite{Mercuri S.,merc_rand, Obregon-OrtegaCruz-Sabido}.

In the 70's MacDowell and Mansouri
\cite{MacDowell-Mansouri}, proposed an action for general relativity,  
based on a $SO(4,1)$ or $SO(3,2)$ gauge group, depending on the sign of the cosmological constant. For a dynamical theory it was necessary to
break the symmetry explicitly, {to obtain the
HP action} plus the Euler class and the cosmological constant.
Some attempts have been made trying to maintain the
full symmetry of the gauge group {and obtain a dynamical theory}. One of them was proposed by
Stelle and West \cite{Stelle-West}, {where} they introduced an auxiliary
vector field $v$, which makes the action invariant under the full
gauge group, but at some point the symmetry is broken by choosing
a preferred direction of $v$. {A different approach that maintains the full symmetry was proposed by Mercuri and Randono
\cite{merc_rand}.  They start with a gauge theory for the de Sitter group and assume a preferred vacuum. This  gives  two spectation values 
that are related to the gravitational constant and the Immirzi parameter. Also, they show that the gauge theory action reduces to the MM
action}. {In this construction when they consider} vanishing torsion,
the Barbero-Immirzi parameter \cite{Immirzi} {is related to the $\theta$ term that appears in Yang-Mills theories}. {Although this
approach, permits straightforward }supersymmetric extensions
\cite{Obregon-OrtegaCruz-Sabido}, the full symmetry {of the}
gauge group must be broken, and  if we want to obtain
the well known Immirzi parameter we have to consider zero torsion
manifolds. Also by adding a Gauss-Bonnet term to
AdS gravity\cite{olea} the authors prove that holographic quantities can be
defined in the boundary of AdS and find that the regularized 4D
action is the MM action. As we can see the MM approach to gravity
has the potential to give new insight in to some of the puzzles of
gravity. {The goal of this paper is to construct a theory inspired by MM gravity that includes torsion.}\\ 
The work is arranged as follows, in Section 2
we review MM gravity. We discuss our proposal motivated by  the MM
theory and present the action with torsion in Section 3. Finally, the last section is devoted to final remarks.


\section{MacDowell-Mansouri gravity.}
The MM theory of gravity is a Yang-Mills type gauge theory with a
gauge group $G \supset SO(3, 1)$. {G} is determined by the
sign of the cosmological constant, $SO(4, 1)$ for $\Lambda > 0$ and $SO(3, 2)$ for $\Lambda < 0$. From phenomenological considerations a positive cosmological constant is favoured  and therefore we will restrict to that case but for $\Lambda<0$ the same procedure applies.

We start with  $SO(4,1)$ as our gauge group,
a 4-dimensional oriented smooth manifold  $\mathcal{M}=\mathbb{R}\times \Sigma$. Where $\Sigma$ is
compact and without boundary and $\mathbb{R}$ represents an
evolution parameter and choose a principal $SO(4,1)$-bundle $P$
over $\mathcal{M}$.

Let
$t_{AB}$ be the elements of $\mathfrak{so}(4,1)$ Lie algebra,
where the indices take the values $A,B, \ldots =0,1,2,3,4$ and
satisfy $
[t_{AB},t_{CD}]=f_{ABCD}^{\ \ \ \ \ \ EF} \ t_{EF}=4\ \eta_{AB,[C}^{\
\ \ \ \ \  E }\eta_{D]}^{\ \ F} t_{EF}.$
The Cartan-Killing form $\widetilde \kappa_{ABCD}$ in the adjoint representation is
\begin{equation} \label{Cartan Killing form RA}
\widetilde{\kappa}_{ABCD}=-\frac{1}{12}{\rm Tr}(t_{AB}t_{CD})=f_{ABEF}^{\ \ \ \ \ \ GH}f_{CDGH}
^{\ \ \ \ \ \ EF}=\eta_{AB,CD},
\end{equation}
where $2\eta_{AB,CD}=\eta_{AC}\eta_{BD}-\eta_{AD}\eta_{BC}$. For
our purposes we take the connection $A$ in the fundamental
representation and split the components as
$A_\mu^{ab}=\omega_\mu^{ab}$ and $A_\mu^{4a}=-\frac{1}{\ell}e_\mu^{a}$,
where $a,b=0,1,2,3$. The field strength is
\begin{equation} \label{field strength}
 F^{ab} =R^{ab}-\ \frac{1}{\ell^{2}}\ e^{a}\wedge e^{b},  \quad F^{4a}= -\frac{1}{\ell}\ T^{a},
\end{equation}
where $R^{ab}$ and $T^{a}$ are  the curvature and  torsion of $SO(3,1)$ and $\ell^2=\frac{3}{\Lambda}$.

To construct the action, we start in the adjoint representation and use the Cartan-Killing form Eq.(\ref{Cartan Killing form RA})
\begin{equation}    \label{FF action RA}
S_{FF}\left[ A\right]= \int_{\mathcal{M}}  {\rm Tr} \ F\wedge F=
\int_{\mathcal{M}}\ F^{AB}\wedge F^{CD}\
\widetilde{\kappa}_{ABCD}=\int_{\mathcal{M}} F^{AB}\wedge F_{AB}.
\end{equation}
This action corresponds to the Pontrjagin class of $SO(4,1)$ \cite{Chandia-Zanelli,Wise}  and does not give any dynamical information.

On the other hand, it is possible to calculate the same action
using the orthogonal decomposition of the Cartan-Killing form by
means of the fundamental representation Eq.(\ref{Cartan Killing form
FR2}), then the action reads
\begin{equation} 
 S_{FF}=\int_{\mathcal{M}} F^{AB}\wedge F^{CD} \kappa_{ABCD}=\int_{\mathcal{M}}\  R^{ab}\wedge R_{ab}-\frac{2}{\ell^{2}}\left[
R^{ab}\wedge e_{a}\wedge e_{b}-T^{a}\wedge T_{a} \right],\label{FF action FR}
\end{equation}
where we can identify the Pontrjagin class for $SO(3,1)$ and the
only  closed 4-form invariant under local Lorentz rotations
associated with the torsion of the manifold, the so-called
Nieh-Yan (N.Y.) class, which is given by
\begin{equation}  \label{Nieh-Yan class}
d(e^{a}\wedge T_{a})=T^{a}\wedge T_{a}-R^{ab}\wedge e_{a}\wedge e_{b}.
\end{equation}
Therefore, action Eq.(\ref{FF action FR}) is purely topological, this is
something that we expected because the information that comes from
the action does not depend on the representation. Finally, from Eq.(\ref{FF action RA}) and Eq.(\ref{FF action FR}) we obtain a well-known result: 
\[\rm{``Pontrjagin}_{SO(4,1)}=\rm{Pontrjagin}_{SO(3,1)}+\rm{N.Y.}"\]
MacDowell and Mansouri \cite{MacDowell-Mansouri} observed that in
order to obtain a dynamical action, it is necessary to break the
symmetry explicitly. The proposal is
\begin{equation}\label{MM action}
S_{MM}=\int_{\mathcal{M}} {\rm Tr} \ (i\gamma^{5}
F\wedge F)=\int_{\mathcal{M}}\ F^{AB}\wedge F^{CD}\
\kappa^{(1)}_{ABCD},
\end{equation}
and from Eq.(\ref{Cartan Killing form so(3,1)}) we get
\begin{equation}
 S_{MM}= \int_{\mathcal{M}} \frac{1}{2}R^{ab}\wedge
R^{cd}\epsilon_{abcd} -\frac{1}{\ell^{2}}R^{ab}\wedge e^{c}\wedge
e^{d}\epsilon_{abcd}+\frac{1}{2\ell^{4}} e^{a}\wedge e^{b}\wedge
e^{c}\wedge e^{d}\epsilon_{abcd},\label{27}
\end{equation}
where we identify the Euler class,   the Palatini action and the
cosmological constant term. From this action, we get the zero torsion condition that allows the
spin connection to be written in terms of the tetrad field  and when substituted in the equation of motion, we
arrive to the Einstein's equations with cosmological constant.

In the background independent  approaches to gravity, the starting
point is the Holst action\cite{Holst}, which is written as  a sum
of the Palatini's action plus the Holst term. An alternative to
introducing the Holst action is to consider the approach given by
Mercuri and Randono \cite{merc_rand}. {Starting with an $SO(4,1)$ gauge invariant action, after symmetry breaking, the
MM action with a topological $\theta$-term arises}
\begin{equation}
S=S_{MM}+\theta S_{FF},\label{randono}
\end{equation}
so if we consider the equation of motion  $T^{a}=0$
we obtain the additional Pontryagin class plus the Holst term with the
Immirzi parameter $\gamma$  related to the $\theta$-term.\\
When non-vanishing torsion is present, Chand\'ia and
Zanelly \cite{Chandia-Zanelli} proposed a torsional contributions to
the chiral anomaly in the form of a N.Y. term. Because of the importance of the Holst term and the N.Y. class, we are
interested in an action where these two terms are present as
independent components.
Using a different Cartan-Killing form $\kappa^{(2)}_{ABCD}$,  Eq.(\ref{Cartan killing form so(3,1) 2}), we construct the action
\begin{equation}
S_{PH}=\int_{\mathcal{M}} F^{AB}\wedge
F^{CD}\ \kappa^{(2)}_{ABCD}=\int_{\mathcal{M}} -R^{ab}\wedge R_{ab}+\frac{2}{\ell^{2}}
R^{ab}\wedge e_{a}\wedge e_{b},\label{PH Action}
\end{equation}
that is the sum of the Pontrjagin class and the Holst
term.  In order to consider the most general action we add Eq. (\ref{randono}) to Eq. (\ref{PH Action}) and also
non-vanishing torsion term. For the MM
approach, it is necessary to consider a linear combination of the
three different actions that could be constructed in the theory
\begin{equation}   \label{General Action}
S_{G}=\mu\ S_{FF}+\nu\ S_{MM}+\rho\ S_{PH},
\end{equation}
where $\mu, \nu, \rho$ are arbitrary constants. Then the action reads
\begin{eqnarray}\label{action_11}
S_{G}&=&\int_{\mathcal{M}} -\frac{\nu}{\ell^{2}}\epsilon_{abcd}R^{ab}\wedge
\ e^{c}\wedge e^{d}+\frac{\nu}{2\ell^{4}}\ \epsilon_{abcd}\ e^{a} \wedge e^{b} \wedge e^{c} \wedge e^{d}+\frac{2\rho}{\ell^2}
R^{ab}\wedge e_{a}\wedge e_{b} \nonumber\\
&+&\frac{\nu}{2} \epsilon_{abcd}R^{ab}\wedge R^{cd}+(\mu-\rho)R^{ab}\wedge
R_{ab}+\frac{2\mu}{\ell^{2}} \left(T^{a}\wedge T_{a}-R^{ab}\wedge
e_{a}\wedge e_{b}\right),\end{eqnarray}
{where the Immirzi parameter is associated with the inverse of
$\rho$.}
This  action contains all the topological
invariants in 4-dimensions, the Euler class, the Pontrjagin class,
the N.Y. class as well as the dynamical part. {Finally we can see that Eq.(\ref{action_11})  matches the deformed SO(4,1) BF model\cite{FS_BF}}.

The equations of motion of
the $S_{G}$ action are
\begin{gather}
D\left(\epsilon_{abcd}e^{c}\wedge e^{d}-\frac{2\rho}{\nu}\
e_{a}\wedge e_{b} \right)=0,\nonumber\\
-R^{ab}\wedge e^{c}\epsilon_{abcd}-\frac{2\rho}{\nu}R^{\
b}_{d}\wedge e_{b} +\frac{1}{\ell^2} e^{a} \wedge e^{b} \wedge
e^{c}\epsilon_{abcd}=0,
\end{gather}
these are similar to the equations of motion for
the Holst action with cosmological constant term, and at first
sight, it could be argued the need of the action $S_{PH}$. 
{If we want to include torsion in the theory, we need  to consider the three terms in Eq(\ref{General Action}). Promoting  the Immirzi parameter to be a  field gives contributions to the torsion\cite{Taveras-Yunes,Torres-Gomez-Krasnov,toloza}. In the same manner we can consider the parameters  $\mu,\nu$ and $\rho$ as fields. This generalization, after the variation, will have an impact on the torsion contributions by $T^a\wedge T_a=0$ condition resulting from the new scalar fields. Then the generalization of the parameters  $\mu,\nu$ and $\rho$ to scalar fields will contribute to the torsion and therefore their inclusion is essential.}


\section{A Gravity action with torsion}
The MM action has been used to understand several properties
of gravity theory, this is achieved by using the similarities between MM
and YM theory. Mercury and Randono\cite{merc_rand} were able to
understand the origin of the Barbero-Immirzi parameter by relating it to a
$\theta$-term. This approach was generalized for the SUSY case \cite{Obregon-OrtegaCruz-Sabido},
where the authors start with  MM supergravity, add a $\theta$-term and  identify the supergravity action, plus the supersymmetric Holst action \cite{kaul}. In
recent works\cite{Torres-Gomez-Krasnov,Taveras-Yunes}, the Immirzi parameter $\gamma$ has been considered to
be a field, therefore this idea
look like a nice starting point to generalize
the MM theory.

In order to generalize the action Eq.(\ref{General Action}), let
us consider the arbitrary constants as fields, $\mu=\mu(x),
\nu=\nu(x), \rho=\rho(x)$ and $\mu$, $\nu$, $\rho\in
C^{\infty}_{0}(\mathcal{M})$ (where $C^{\infty}_{0}$ denotes the
space of $C^{\infty}$ functions on $\mathcal{M}$ with compact
support in $\mathcal{M}$), then the general action is
\begin{eqnarray}\label{sg}
 S_{G}&=&\int_{\mathcal{M}}\left(\mu(x)-\rho(x)\right)R^{ab}\wedge
R_{ab}+\frac{1}{2}\nu(x)R^{ab}\wedge R^{cd}
\epsilon_{abcd}+\frac{2}{\ell^{2}} \mu(x)D\left(e^{a}\wedge T_{a}\right)\\
 &~&-\frac{1}{\ell^{2}}  \nu(x) R^{ab}\wedge e^{c}\wedge
e^{d}\epsilon_{abcd}+\frac{2}{\ell^{2}} \rho(x) R^{ab}\wedge
e_{a}\wedge e_{b} +\frac{1}{2\ell^{4}} \nu(x)\epsilon_{abcd} e^{a}
\wedge e^{b} \wedge e^{c} \wedge e^{d}.\nonumber
\end{eqnarray}
It is important to note that in this work we will not consider
matter contributions (there is a proposal for an action depending on these fields\cite{Taveras-Yunes}).\\
The equations of motion coming from the action Eq.(\ref{sg}) are
\begin{eqnarray} \label{Equation of motion omega}
\mu(x)& \Rightarrow& R^{ab}\wedge\left[ e_{a}\wedge e_{b}-\frac{\ell^{2}}{2}R_{ab} \right]=T^{a}\wedge T_{a},   \\ 
 \rho(x) &\Rightarrow&  R^{ab}\wedge \left[ e_{a}\wedge e_{b}-\frac{\ell^{2}}{2}R_{ab}  \right]=0, \nonumber \\
 \nu(x) &\Rightarrow& \frac{\ell^{2}}{2} R^{ab}\wedge R^{cd}\epsilon_{abcd}=\left[R^{ab}\wedge e^{c} \wedge e^{d}-\frac{1}{2\ell^{2}}e^{a}\wedge e^{b}\wedge e^{c}\wedge e^{d}\right]\epsilon_{abcd}, \nonumber  \\
\omega(x)& \Rightarrow& D\left(e_{a}\wedge e_{b}\right)=-\frac{2\rho(x)\eta_{ab}^{\ \ cd}+\nu(x)\epsilon_{ab}^{\ \  cd}}{4\rho^{2}(x)+4\nu^{2}(x)}\bigg( 2D\mu(x)\wedge [\ell^{2}R_{cd}+e_{c}\wedge e_{d}]\nonumber \\
 &~&-2D\rho(x)\wedge[\ell^{2}R_{cd}-e_{c}\wedge e_{d}]+D\nu(x)\wedge [\ell^{2}R^{mn}-e^{m}\wedge e^{n}]\epsilon_{cdmn}\bigg),  \nonumber \\
e(x)& \Rightarrow &R^{ab}\wedge
e^{c}\epsilon_{abcd}-\frac{1}{\ell^{2}} e^{a}\wedge e^{b}\wedge
e^{c}\epsilon_{abcd}=\frac{2}{\nu(x)} D\mu(x)\wedge
T_{d}-\frac{2\rho(x)}{\nu(x)} DT_{d},\nonumber
\end{eqnarray}
to write the equation of motion  for $\omega$ we have used
the projector $\mathcal{A}$ defined as follows $
\mathcal{A}_{abcd}=\alpha \ \eta_{ab,cd}+\beta \ \epsilon_{abcd} $
where $\alpha, \beta$ are fields such that
$\alpha^{2}\neq-4\beta^{2}$, then  its inverse is given by $
(\mathcal{A}^{-1})_{abcd}=\dfrac{1}{\alpha^{2}+4\beta^{2}}\left(
\alpha\ \eta_{ab,cd}-\beta \epsilon_{abcd} \right)$ such that
$\mathcal{A}\mathcal{A}^{-1}=\mathcal{A}^{-1}\mathcal{A}=\eta_{ab,cd}$.\\
Let us analyze the equations of motion in order to obtain physical
implications given by the new fields. We will pointing out some
implications given by the equations of motion
\begin{itemize}
\item From  the equations for $\mu$
and $\rho$ in (\ref{Equation of motion omega}), we obtain
\begin{equation}\label{torsion eq}
T^{a}\wedge T_{a}=0 \qquad \textrm{that is equivalent to} \qquad
R^{ab}\wedge e_{a}\wedge e_{b}=d(e^{a}\wedge T_{a}).
\end{equation}
Then the Holst term is a topological object in this theory.
\item From the same equations of motion we obtain $R^{ab}\wedge e_{a}\wedge e_{b}=\frac{l^{2}}{2}\ R^{ab}\wedge
R_{ab}$. This equation has the following two implications. The first can be obtained once we integrate over the manifold,
\begin{equation}   \label{Pontryagin equals Holst}
 \int_{\mathcal{M}} R^{ab}\wedge e_{a}\wedge e_{b}=\frac{\ell^{2}}{2} \int_{\mathcal{M}} R^{ab}\wedge
R_{ab},
\end{equation}
then the Holst term is equal, on shell,  to the Pontrjagin class
of $SO(3,1)$. This result was obtain by Liko \cite{Liko} by means
of the equation of motion for the tetrad field and reenforces the
topological implication over it. In our case we have recovered
that result only by considering equation of motions for the new
fields in a non vanishing torsion scheme. The second
implication is given by means of Eq.(\ref{Equation of motion omega}), as
follows
\begin{equation}
d\bigg(e^{a}\wedge T_{a}+\frac{l^{2}}{2}\ \Big(\omega^{ab}\wedge
d\omega_{ab}+\frac{3}{2}\ \omega^{ab}\wedge \omega_{a}^{\ c}\wedge
\omega_{cb}\Big) \bigg)=0,
\end{equation}
then the torsion term and the Chern-Simons class are related modulo a
closed three form, so there is a cohomology class relation between
them
\begin{equation}
e^{a}\wedge T_{a}+\frac{\ell^{2}}{2}\ \Big(\omega^{ab}\wedge
d\omega_{ab}+\frac{3}{2}\ \omega^{ab}\wedge \omega_{a}^{\ c}\wedge
\omega_{cb}\Big) =d\chi,\label{cohomology}
\end{equation}
where $\chi$ is a two-form field without any constraint over it.
Then, from the last equation and the equation of motion for
$\omega$, we obtain that in general the field $\chi$ is given by
\begin{equation}
\chi=\chi(\mu, \nu, \rho, e,\omega).
\end{equation}
An important observation is that if we consider the
theory in three dimensions, this closed three form disappears.\\
\item From the equation of motion of $\nu$ and $e$ in (\ref{Equation of motion omega}), we get
\begin{equation} \label{Equation of motion nu and e}
 \frac{\ell^{2}}{2}R^{ab}\wedge R^{cd}
\ \epsilon_{abcd}-\frac{1}{2\ell^{2}} e^{a}\wedge e^{b}\wedge
e^{c}\wedge
 e^{d}\ \epsilon_{abcd}=\frac{2}{\nu(x)}\ d\mu(x)\wedge T_{d}\wedge
 e^{d}-\frac{2\rho(x)}{\nu(x)}d(T_{d}\wedge
 e^{d}).
\end{equation}
This equation relates the cosmological term and Euler class to
the torsion contribution, which suggest that the torsion produced
by the appearance of these new fields is given by  topological
effects.

 \item Finally if we consider the equation of motion of
$\omega$, multiply both sides by $e^{n}$ and by using the equation
of motion for the tetrad field, we find
\begin{equation}  \label{Torsion second fundamental derivation}
 e\ \Phi^{1}_{m}d^4x=\Phi^{2}\wedge DT_{m}+\Phi^{3}\wedge
 T_{m}+e_{m}\wedge e^{n}\wedge T_{n},
\end{equation} 
where $e$ is the non-vanishing determinant of the tetrad and
\begin{eqnarray}
\Phi^{1} &=& -\frac{\nu(x)}{4}\ \partial_{m} \mu(x), \nonumber\\
\Phi^{2}&=&\ell^{2}
\left( \rho(x) \left( \frac{\rho(x)}{\nu(x)}+1 \right) d\nu(x) \right), \nonumber\\
\Phi^{3} &=&-\ell^{2} \left(  \frac{\rho(x)}{\nu(x)}\ d\nu(x) +\nu(x)
d\mu(x)-d\rho(x) \right)\wedge d\mu(x).
\end{eqnarray}
\end{itemize}
{Let us start by considering $T_{a}\wedge e^{a}=0$, then from Eq.(\ref{cohomology}) we have}
\begin{equation}
\omega^{ab}\wedge d\omega_{ab}+\frac{3}{2}\ \omega^{ab}\wedge
\omega_{a}^{\ c}\wedge \omega_{cb} = d\chi,
\end{equation}
{unfortunately, solving for $\chi$ is a difficult ordeal. A method\footnote{It was presented in the context of non abelian fluid mechanics.}  for solving Eq.(23) for semi compact groups was developed by Jackiw\cite{Jackiw} et.al., as we are working with $SO(4,1)$ we can not use it. Although a generalization of Jackiw's approach for our purposes is interesting, it is beyond the scope of this paper.}

From the condition $e_{a}\wedge T^{a}=0$, we can consider two
different solutions for the torsion. First let us consider the case when the torsion vanishes and its
covariant derivative, i.e. $T^{a}=0$ and $DT^{a}=0$, leading to
$\omega=\omega(e)$, and, when we substitute back into the
torsional modified Einstein's equation,  we have the usual
equation in vacuum. Then from equation (\ref{Torsion second
fundamental derivation}) we find
\begin{equation}
\nu(x)\ \partial_{m}\mu(x)=0,
\end{equation}
this equation has two solutions, $\nu=0$ for all $x\in \mathcal{M} $ and
$\mu$ is a nonzero constant. The first gives rise to trivial
implications as we can observe from the equations of motion, so we
consider the second case. From the equation of motion for $\omega$
we find
\begin{equation}
(-2D\rho(x)\eta_{ab,cd}+D\nu(x)\epsilon_{abcd})\wedge \left[\ell^{2}
R^{cd}-e^{c}\wedge e^{d} \right]=0,
\end{equation}
but the last equations involves the two independent
$\mathfrak{so}(3,1)$ metric forms, each one multiplied by  an
independent field, so it implies that $\nu, \rho$ are constants
too. Also we recover the usual Einstein's  equations of motion. 

The second family solutions are given by
\begin{equation} \label{General Torsion}
T^{a}=\beta \wedge e^{a},
\end{equation}
where $\beta$ is a arbitrary 1-form. Therefore, we have two
solutions for the spin connection
\begin{equation}  \label{Equation omega resume}
\omega = \omega(\mu, \nu, \rho, e)\ \ \textrm{and}  \ \ \omega =
\omega( \beta,  e),
\end{equation}
If we consider as fundamental fields those appearing in the
action, then we have to find a relation among $\mu, \nu, \rho$ and
$ \beta$, i.e., find solutions of the form $\beta=\beta(\mu, \nu,
\rho)$. To find this relation, let us rewrite equation
(\ref{Torsion second fundamental derivation}) as
\begin{equation}
\Phi^{1}_{m}=\mathcal{D}_{m}^{\ \ c} \beta_{c},
\end{equation}
where we have defined the differential operator $\mathcal{D}$
as
\begin{equation}
\mathcal{D}_{m}^{\ c}= -\epsilon_{m}^{\ \ abc}
\Phi^{2}_{a}\partial_{b}-\epsilon_{m}^{\ \ abc}\Phi^{3}_{ab},
\end{equation}
and  $\beta$ is written as
\begin{equation} \label{Beta formal solution}
\beta_{c}=\beta_{c}^{0}+\int G_{c}^{\ \
b}(x,x')\Phi^{1}_{b}(x')d^4x',
\end{equation}
where $G_{a}^{\ \ b}$ is the Green function and $\beta_{c}^{0}$ is
the  solution for the homogeneous equation\cite{Mukhanov-Winitzki,Grubb-book}. We observe that it
is possible to find $\beta=\beta(\mu, \nu, \rho)$ so the torsion
that we have considered is consistent. Then all the
information of the fields $\mu,\nu,\rho$ is encoded in
$\beta$. To find the connection as a function of the tetrad
and $\beta$,  let us consider the equation
(\ref{General Torsion}) and let us define
$\mathcal{A}^{ab}=\omega^{ab}- \beta \delta^{ab}$, then the
torsion equation reads 
\begin{equation}
de^{a}+\mathcal{A}^{a}_{\ b}\wedge e^{b}=0,
\end{equation}
the last equation is similar to the usual zero
torsion condition, and in order to find $\omega$, we proceed as
the usual case \cite{PeterPeldan} and get
\begin{equation}
\omega_{\gamma ab}=\frac{1}{2}\ e_{\gamma}^{\ c}\{
\Omega_{cab}+\Omega_{bca}- \Omega_{abc}\}-(e_{\gamma
a}\beta_{b}-e_{\gamma b}\beta_{a}),
\end{equation}
where $\Omega_{abc}=e^{\mu}_{\ a}e^{\nu}_{\
b}\partial_{[\mu}e_{\nu]}^{\ \ c}$, then as we can observe if
$\beta=0$, we recover the usual case. Finally the spin connection
is written as
\begin{equation}
\omega=\omega(e)+\omega(\beta(\mu, \nu, \rho)),
\end{equation}
and Einstein's equations are modified as follows
\begin{equation}
R^{ab}(\omega(e))\wedge e^{c}\epsilon_{abcd}-\Lambda e^{a}\wedge
e^{b}\wedge e^{c}\epsilon_{abcd}= \Xi_{d}(\beta(\mu, \nu, \rho),e),
\end{equation}
where
\begin{equation}
 \Xi_{d}=R^{ab}(\omega(\beta))\wedge e^{c}\wedge
\epsilon_{abcd}+R^{ab}(\omega(\beta),\omega(e))\wedge
e^{c}\epsilon_{abcd}+\frac{2}{\nu}d\mu\wedge \beta\wedge
e_{d}-\frac{2\rho}{\nu} d\beta\wedge e_{d},
\end{equation}
as the energy-momentum current of matter associated to the
presence of the new fields coupled to the original action. The
dynamical behavior of these fields could be computed once we have
the explicit solution for the $\beta$ field from (\ref{Beta formal
solution}), and by using the Bianchi's identity in the Einstein's
equation of motion.


\section{Final Remarks}

{In this paper inspired by the MM theory of gravity, we have written a generalization, by replacing the coupling constants with fields, on the different
invariants that can be {constructed} in 4 dimensions.
By analyzing the equations of motion for the new fields, we see that we have Einstein equations with torsion contributions.}
We consider two different Cartan-Killing forms that can be
derived from the Lie algebra in $\mathfrak{so}(4,1)$, one in the
adjoint representation and the other one coming from the
fundamental representation. We use a representation of the Lie algebra as a direct sum of the two vector spaces $
\mathfrak{so}(4,1)\cong \mathfrak{so}(3,1)\oplus \mathbb{R}^{3,1}$ and the actions constructed are topological. The dynamics is obtained by explicitly breaking $\mathfrak{so}(4,1)$ to $\mathfrak{so}(3,1)$. In
$\mathfrak{so}(3,1)$, it is possible to find two Cartan-Killing
forms $\epsilon^{abcd}$ and $\eta_{ab,cd}$, and identify two
metrics coming from the broken sector and one related to the
unbroken one. In MM models, one usually works with the form
$\epsilon^{abcd}$, but in order to have a general contribution to the
dynamics, we constructed the action from a linear combination of
the Cartan-Killing forms. We obtain the Palatini action, the
cosmological constant term, the Euler and Pontrjagin terms (as in
MM) but also get the Nieh-Yan class independent from Holst term.

The introduction of three arbitrary parameters (one of them related to the Immirzi's
parameter), is inspired by works treating the Immirzi parameter as a field
\cite{Torres-Gomez-Krasnov,Taveras-Yunes}, we considered the new
parameters as fields and calculated the dynamics coming from
these new fields. As expected, we get a non zero torsion {theory}, that in
general depends on the new fields.
Also, we find  that the Holst {term} is related to the Pontrjagin class of $SO(3,1)$.

The presence of the new parameters might induce an effective non constant $\Lambda$. This new  $\Lambda$ can depend on the new fields and the volume of the universe and therefore might shed some light 
on the smallness of the cosmological constant. This issue is under research and will be reported elsewhere.

\section*{Acknowledgments}
This work is supported  by CONACyT  research grants 167335, 257919, 258982. J.C.L-D is supported by UAZ-2017 grant. J.E.R.Q is supported by CONACyT postdoctoral grant. M. S is supported by  DAIP1107/2016 and by the CONACyT program  ``Estancias sab\'aticas en el extranjero'',   grant 31065.

\appendix
\section{}
As pointed by Wise\cite{Wise}, in the description of rolling
geometries \cite{Helgason-book,Sharpe-book}, the spacetime
geometries relevant to gravity are of a special type called
reductive geometry. In particular, $SO(4,1)$ is a reductive
geometry where 
\begin{equation}    \label{Orthogonal splitting}
\mathfrak{so}(4,1)\cong \mathfrak{so}(3,1) \oplus
\mathfrak{so}(4,1)/\mathfrak{so}(3,1).
\end{equation}
To visualize this splitting we
consider the fundamental representation of the de Sitter Lie
algebra in the basis
\begin{equation}
\left\{  -\frac{1}{2}\ \gamma^{[a} \gamma^{b]}, \frac{1}{2}\
\gamma^{5}\gamma^{c} \right\},
\end{equation}
where we have adopted the complex $4\times4$ matrix representation
of the Clifford algebra
\begin{equation}
 \gamma^{a}\gamma^{b}+\gamma^{b}\gamma^{a}=2\eta^{ab},
\end{equation}
and as usual, we take
\begin{equation}
 \gamma^{5}=\frac{i}{4!}\epsilon_{abcd}\gamma^{a}
\gamma^{b}\gamma^{c} \gamma^{d}=i\gamma^{0} \gamma^{1}\gamma^{2}
\gamma^{3}.
\end{equation}
In the fundamental representation the generators are given by
\begin{equation}
t_{AB}=\left(
\begin{array}{cc}
  t_{ab} & t_{a4} \\
  t_{4b} & 0
\end{array}\right)
=\frac{1}{2 (4^{1/3})}\left(
\begin{array}{cc}
  - \gamma_{[a} \gamma_{b]} &  \gamma_{b}\gamma_{5} \\
    \gamma_{5}\gamma_{b}    & 0
\end{array}\right).
\end{equation}
and satisfy the Lie algebra
\begin{eqnarray}
[t_{ab},t_{cd}] &=& -\frac{1}{2} \left(\eta_{[a|c|}\eta_{b]}^{\ f}\eta_{d}^{\ g}-\eta_{[a|d|}\eta_{b]}^{\  f}\eta_{c}^{\ g}
+f\leftrightarrow g\right)t_{fg}, \nonumber\\
{[t_{ab},t_{c4}]} &=& -\frac{1}{2} \eta_{ab,c}^{\ \ \ \ d}t_{d4},\\
{[t_{a4},t_{b4}]} &=& -\frac{1}{2} \eta_{ab}^{\ \  cd}t_{cd}.\nonumber
\end{eqnarray}
The corresponding Cartan-Killing form is calculated
straightforward
\begin{equation}  \label{Cartan Killing form FR2}
\kappa_{ABCD}=-\frac{2}{4^{2/3}}\ Tr(t_{AB}t_{CD})=
\left(\begin{array}{cc}
\eta_{ab,cd} & 0 \\
 0 & \kappa_{mn}
\end{array}\right),
\end{equation}
where
\begin{equation}
\kappa_{mn}=\left(
\begin{array}{cc}
   \kappa_{m4n4} &  \kappa_{m44n}\\
   \kappa_{4mn4} &  \kappa_{4m4n} \\
\end{array}
\right)=\frac{1}{2}\ \left(
\begin{array}{cc}
    \eta_{mn} & - \eta_{mn}\\
  - \eta_{mn} &   \eta_{mn} \\
\end{array}
\right).
\end{equation}
Now the orthogonal splitting invariant under $SO(3,1)$, of the
Cartan-Killing form is  evident and
\begin{equation} \label{Cartan killing form splitting}
\mathfrak{so}(4,1)\cong \mathfrak{so}(3,1)\oplus \mathbb{R}^{3,1},
\end{equation}
as vector spaces instead of Lie algebras\cite{Wise}.\\
From Eq.(\ref{Cartan Killing form FR2}), we recognize
$\eta_{ab,cd}$ as the Cartan-Killing form for
$\mathfrak{so}(3,1)$, but in  four dimensions we have another
invariant form,  the Levi-Civita tensor $\epsilon_{abcd}$, and in
the de Sitter algebra is possible to obtain this form by means of
$\gamma^{5}$,
\begin{equation}   \label{Cartan Killing form so(3,1)}
\kappa_{ABCD}^{(1)}=-\frac{2}{4^{2/3}}
{\rm Tr}(i\gamma^{5}t_{AB}t_{CD})=\frac{1}{2}
\left(\begin{array}{cc}
\epsilon_{abcd} & 0 \\
 0 & 0
\end{array}\right).
\end{equation}
We  observe that the presence of ($i\gamma^{5}$) in the trace
term, breaks the symmetry from $\mathfrak{so}(4,1)$ down to
$\mathfrak{so}(3,1)$. The invariant form constructed in
Eq.(\ref{Cartan Killing form so(3,1)}) is often used in the MM
formulation of gravity \cite{Stelle-West}.\\
We can recover the form $\eta_{ab,cd}$ by means of
\begin{equation}   \label{Cartan killing form so(3,1) 2}
\kappa^{(2)}_{ABCD}=-\frac{2}{4^{2/3}}
{\rm Tr}\left((i\gamma^{5}t_{AB})(i\gamma^{5}t_{CD})\right)=
-\frac{1}{2}\left(
\begin{array}{cc}
 \eta_{ab,cd} & 0 \\
 0  & 0
\end{array}\right).
\end{equation}
The last $\mathfrak{so}(3,1)$-invariant form isn't usually
presented in the literature, but it is very important in the
theory of MM.


\end{document}